# Terahertz emission from silicon carbide nanostructures

*N.T. Bagraev*[1,2], *S.A. Kukushkin*[1], *A.V. Osipov*[1], *L.E. Klyachkin*[2], *A.M. Malyarenko*[2], *V.S. Khromov*[1,2]

[1] Institute of Problems of Mechanical Engineering, Russian Academy of Sciences, 199178 St. Petersburg, Russia
[2] Ioffe Institute, 194021 St. Petersburg, Russia

E-mail: bagraev@mail.ioffe.ru

For the first time, electroluminescence detected in the middle and far infrared ranges from silicon carbide nanostructures on silicon, obtained in the framework of the Hall geometry. Silicon carbide on silicon was grown by the method of substitution of atoms on silicon. The electroluminescence from the edge channels of nanostructures is induced due to the longitudinal drain- source current. The electroluminescence spectra obtained in the terahertz frequency range, 3.4, 0.12 THz, arise due to the quantum Faraday effect. Within the framework of the proposed model, the longitudinal current induces a change in the number of magnetic flux quanta in the edge channels, which leads to the appearance of a generation current in the edge channel and, accordingly, to terahertz radiation.

**Keywords:** silicon carbide on silicon, terahertz emission, electroluminescence, nanostructure, quantum Faraday effect.

## 1. Introduction

Recently, close attention has been paid to studies aimed at producing and using the terahertz radiation (THz radiation) — the region of the electromagnetic wave spectrum with the frequencies within the range $0.1-10$ THz (1 THz = 1012 Hz, this frequency corresponds to the wavelength of 300 $\mu$m and the energy of 4.14 meV) [1].

Similarity radiation has a number of unique characteristics. First, unlike X-rays, the energy of THz radiation is low and causes no molecule ionization. Thus, there is no risk of irradiation/damage to the material [1]. In practice, it can be used for nondestructive testing of fine structures, such as polymer, composite, paint & varnish and protective materials, pharmaceutical tablet coatings [2]. It is noted that it is possible to investigate by THz-radiation the phonon spectrum of nonorganic and organic molecular crystals, as well as hydrogen bond networks of organic molecular crystals. The opacity of conducting materials for THz radiation enables to distinguish conducting and non-conducting phases with high contrast in the THz-spectrum [1].

Secondly, the energies of vibrational and rotational transitions in the molecules and macromolecules are present in THz frequency region thereby making THz radiation a powerful tool for their identification and study of their structure. This terahertz spectrometer-detector can be adapted for investigating DNA sequences [1,3], revealing tumor-specific proteins, registrating changes in a molecular structures in a tumor-affected tissue in order to reveal and investigate the behavioral dynamics of the latter.

Thirdly, the polar molecules (water and body fluids) strongly absorb the THz radiation. Thus, it is now possible to use the THz radiation for biology and medicine for high-resolution diagnostics, which is based in recording the variations in water content in the tissues or the fluid composition itself. With its penetrability through media (opaque for visible and near-infrared radiation), it is noted that it is possible to apply the THz radiation to track a skin hydration level in real time [1], as well as for early diagnostics of skin cancer [1].

Initially, the sources of microwave high-frequency radiation (i.e. with frequency close to the low THz range) included vacuum electronics devices [4, 5]: klystrons, travelling-wave tubes, gyrotrons, backward-wave tubes [6]. The latter can cover the frequency range to 1.2 THz. In order to obtain the coherent THz radiation, several techniques are applied today: optical pumping of the active medium (gas) with external laser radiation, frequency down conversion using two lasers with a difference in the operating frequencies that is a required value of several THz [5]. These methods require cumbersome and expensive equipment. In particular, the study [7] applies a commercial system of the THz-radiation source detector (produced by the Menlo Systems) with the femtosecond laser. The total weight of this device (the optical & mechanical part and the control unit) can be 54 kg [8]. For the laser technique, the THz-range radiation was demonstrated using the quantum cascade laser in 2002 [9]. It was the next step after demonstration of the mid-infrared range radiation on the cascade laser in 1994 [10]. However, all the above-listed is quite detectors and there is still a problem of creating compact sources and receivers of THz radiation. One of the first attempts was to produce THz-radiation in the structure containing Josephson HTS contacts [11]. It has confirmed producibility of the systems of compact sources and receivers of THz radiation based on solid-state samples, wherein macroscopic effects of quantum interference exhibit. The next step was creation of similar compact sources designed to operate at the room temperature. These sources were based on silicon nanosandwiches and used in practical medicine both in diagnostics and treatment of socially significant diseases [12]. Radiation is generated in these nanosandwiches in edge channels confined by dipole boron centers that form networks of Josephson junctions [13]. The obtained results have shown that in order to increase the power of THz radiation, it is advisable to use wide-band materials [14] for nanosandwich basis. Silicon carbide is such a material among the semiconductor compounds.

## 2. Methods

As the elements of this system can be utilized semiconductor samples, which contain thin films of monocrystalline SiC grown on the surfaces (100), (110) and (111) of monocrystalline Si. The studies [15–17] have developed the method of producing these films in the method of matched atom substitution by the chemical reaction of silicon with gaseous carbon monoxide (CO). The detailed description of the processes during the SiC growth by the atom substitution method and its synthesis technology can be found in the reviews [18, 19].

The principle difference of this SiC films growth mechanism from their growth in other methods is that the structure of the initial cubic Si lattice is kept, thereby ensuring the growth of a cubic 3C-SiC polytype [18–20]. It was confirmed by the electron-microscopic studies, which have also shown no mismatch dislocation of the lattices on the 3$C$-SiC(111)/Si(111) interface; there are stacking faults with layers of hexagonal phases instead of them at the interphase interface [21]. The term "matched" means that removal processes of the Si atom out of the lattice and the displacement of the C atom into its position due to the reaction

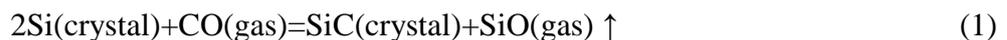

$$2Si(crystal)+CO(gas)=SiC(crystal)+SiO(gas) \uparrow \qquad (1)$$

are simultaneously [20]. The epitaxy of the SiC films on Si due to matched substitution of the half of Si atoms to C atoms with no lattice mismatch dislocation ensures high crystalline perfection of the SiC films [18–21].

The reaction (1) of silicon carbide synthesis is a two-stage process. The first one forms silicon vacancy-interstitial carbon atom complexes. The second one displaces the carbon atoms towards the silicon vacancies by forming silicon carbide. The activated complexes transform into silicon carbide, while generated vacancies assemble in pores below the silicon carbide layer. As a result, the silicon carbide film forms to be partially hanging over the silicon pores. Therefore the films produced by this method have no elastic stress [15–21]. Thus the film is oriented by an "old" crystal structure of the initial Si matrix not just a substrate surface as it is usually implemented in traditional film growth procedures.

An important feature of this SiC growth method includes appearance of the interface layer of thickness of several nanometers on the SiC/Si interface, which has non-standard optical and electrophysical properties, which are caused by shrinkage of the initial lattice. At the last stage of turning silicon into silicon carbide the Si material initial lattice with the lattice parameter of 0.543 nm "collapses to" the SiC cubic lattice with the parameter of 0.435 nm; this shrinkage process is happening within the substrate plane [18,21].

At the same time, a new phase, i.e. the silicon carbide, separates from the silicon matrix and puts on the latter extremely high pressure exceeding 100GPa. It would be impossible to produce SiC with such a good structure under such high pressures if every fifth lattice cell of silicon carbide did not align accurately with every fourth cell of silicon. This shrinkage results in matched arrangement of chemical bonds: every fifth SiC bond with every fourth Sic bond. The other bonds either get disrupted (thus producing vacancies and pores), or are subjected to compression, which alters the structure of surface bonds of SiC adjacent to Si and leads to its transformation into a "semimetal". This effect has been observed for the first time experimentally using the spectral ellipsometry technique in the range of photon energies 0.5−9.3 eV [22].

The quantum-chemical calculations of the paper [22] showed that in the process of dislocation-free matching of SiC and Si lattices that differ by 20%, the SiC film with its Si surface facing the substrate attracts one of the 16 Si atoms in the proximal double layer of the substrate atoms. At the same time, 22 of the 25 Si atoms form chemical bonds with the Si substrate atoms, while 3 atoms of 25 (i. e. 12%) do not form bonds as they are far way from the substrate atoms (more 3 °A); $p$-electrons of these Si atoms in SiC mainly contribute to a narrow and sharp peak of the density of electron states 3$C$-SiC(111)/Si(111) close to the Fermi energy. In other words, the 3$C$-SiC(111)/Si(111) interface should have unusual electrophysical properties; specifically it should be a good electric conductor.

This interface also have certain unusual magnetic properties. The study [23] has investigated a sample containing the silicon carbide on silicon film (110), which was doped with boron after film formation in conditions of non-equilibrium diffusion from the gas phase in an excessive flux of silicon vacancies from the surface of the silicon carbide on silicon samples produced. The technological parameters of the sample (F5) are specified in [23]. The "dia-para hysteresis" of magnetic susceptibility which is known as experimental demonstration of the Meissner–Ochsenfeld effect and oscillations of magnetic susceptibility with the periods 2, 13 and 300Oe were obtained which identify implementation of quantum interference conditions within the sample plane in the vicinity of microdefects. The investigation of these oscillations exhibit the de Haas–van Alphen (DHVA) and Aharonov–Bohm (AB) effects, which are associated with quantization of the momentum and quantization of the magnetic flux at the room temperature, respectively. Thus, it has demonstrated effective suppression of the electron-electron interaction. It is made possible by the formation of dipole boron centers with a negative correlation energy, that confine the edge channels of the studied structure and, respectively, govern the characteristics of the DHVA and AB oscillations [24]. It should be noted that edge channels may be formed not only by dipole boron centers, but also by dipole centers of the "silicon vacancy – interstitial carbon atom" type, which are always present in SiC/Si structures grown by matched substitution of atoms on the (111) and (110) silicon substrate surfaces [20]. Respectively, the observation of quantum interference in edge channels is interrelated with the above-mentioned "dia−para-hysteresis" of the magnetic susceptibility. Thus, an important role of the quantum interference regions is determined for formation of Josephson networks inside the edge channels of nanosandwiches.

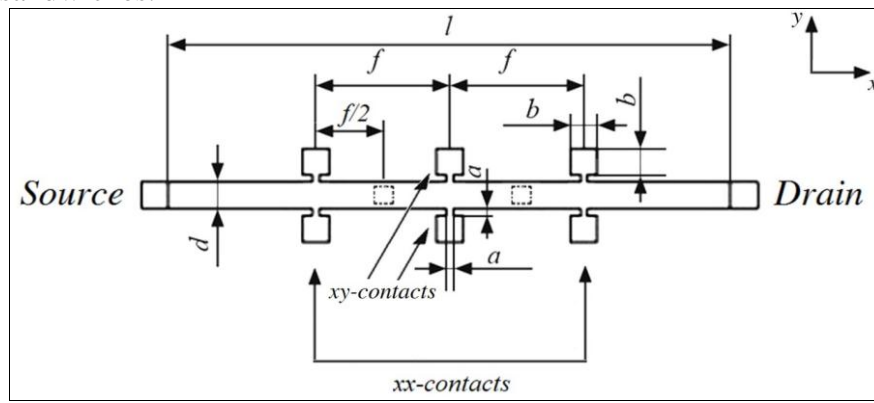

**Figure 1.** Hall geometry of contacts on the surface of the studied structure. Parameters in $\mu$m: $a = 50$, $b = 200$, $d = 200$, $l = 4200$, $f = 1000$. The dashed contours denote the positions of vertical contacts $b \times b$ in size formed above the structure's surface.

## 3. Results and discussion

The Bruker Vertex 70 Fourier spectrometer was used to study the electroluminescence spectrum of the sample provided. When passing the longitudinal current $I_{ds}$ through drain-source contacts within the milliampere range, there was evidently the terahertz radiation spectrum in the near-(Fig. 2, *a*) and far- (Fig. 2, *b* and *c*) IR-ranges. The dips in the spectra are due to radiation absorption by water vapor and carbon dioxide, whose absorption lines positions were identified using the HITRAN database [25]. Thus, the features around 15 $\mu$m (20 THz) are correlated to the $CO_2$ absorption lines, while the features within the range 5−8 $\mu$m (60.0−37.5 THz) — to a set of the water vapor absorption lines.

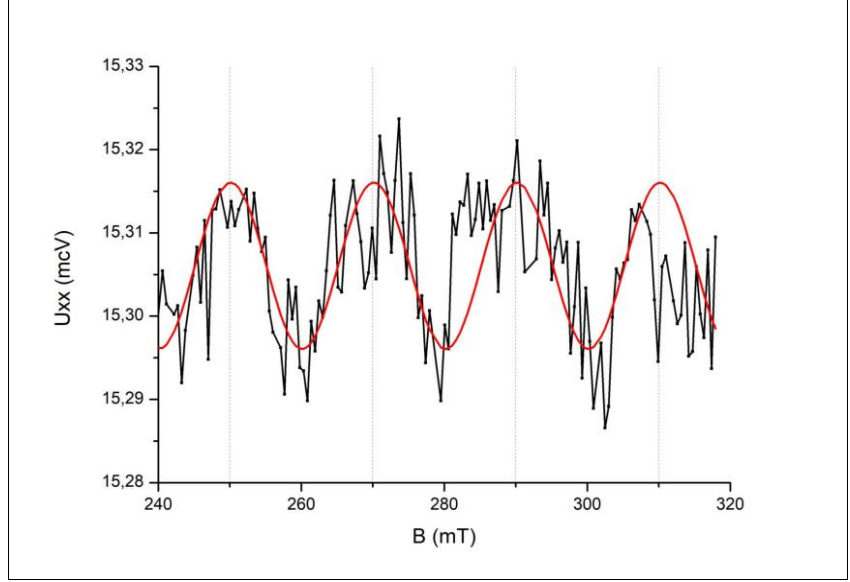

**Figure 2.** Electroluminescence spectra in the mid- (5 — 27 $\mu$m (*a*)) and far-IR-ranges (14 — 333 $\mu$m (*b*), 182 — 667 $\mu$m (*c*)); $I_{ds}$ = 30mA, $T$ = 300K.

The obtained results are explained below within the model of THz radiation generation from the edge channels of the studied structure.
The magnetic field dependences of the longitudinal voltage $U_{xx}$ of the studied sample (Fig. 3) provide an experimental value of the AB oscillations period: $\Delta B$ = 20 mT, from which can be estimated a longitudinal size $l_0$ edge channel region where interference of single carrier occurs

$$l_0 = \frac{\Phi_0}{\Delta B d_0}, \quad (2)$$

where $\Phi_0 = h/e$ — magnetic flux quantum while as the edge channel's width $d0$ the value of 1.54 nm can be considered, which is specified in the paper [18] as a typical distance between shrinkage pores, which form on the surface of the silicon substrate and are covered by the SiC layer. Then, the value obtained is $l_0 \approx 0.1344 \cdot 10^{-3}$ m $\approx 134$ $\mu$m. It can be concluded that the edge channel of the studied structure consists of the regions of interference of the single carriers. Each region ("the pixel") consists of layers containing boron dipoles and having the size $S_{pixel}$ = 134 $\mu$m × 1.54 nm, through which the carrier is tunneling; these two layers confine a SiC core of the height of 2 nm, as the process of sample doping with boron the concentration profile of the latter changes at the distances of about 2 nm in accordance with the measurements of the DHVA oscillations [24]. The scheme of the pixel is shown on Fig.4.

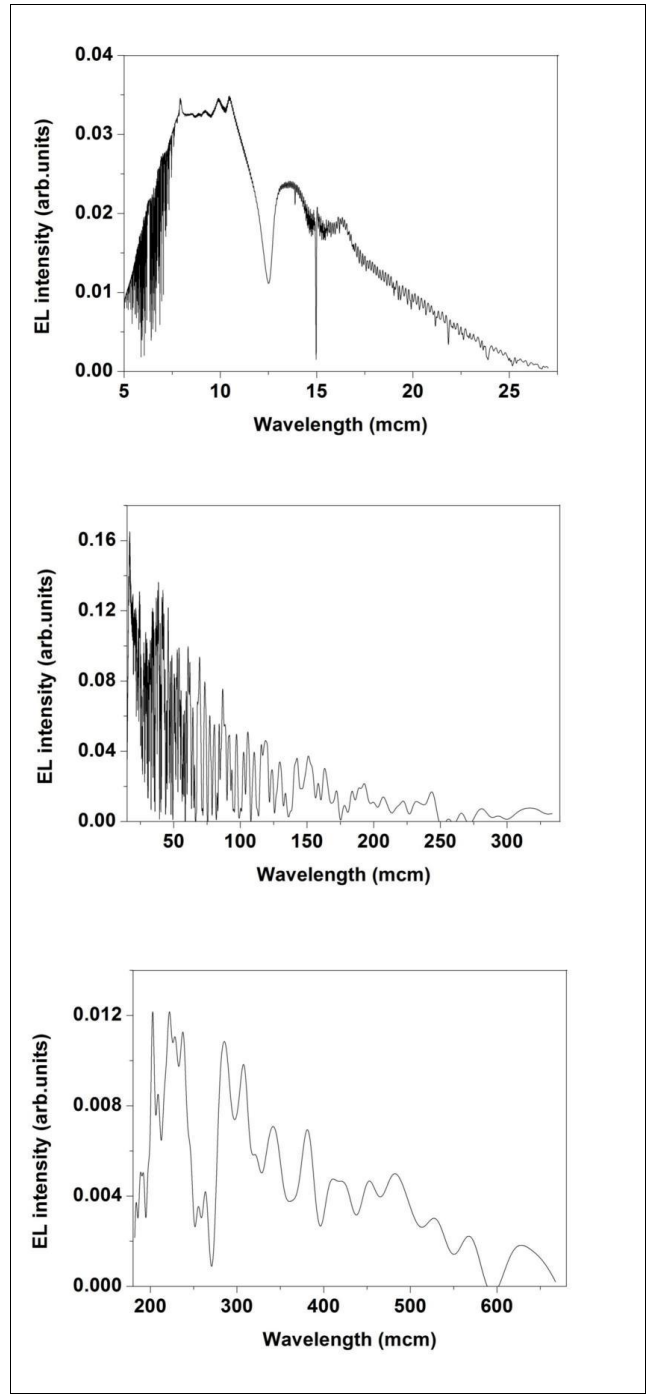

**Figure 3.** Dependence of the longitudinal voltage of the *xx* nanosandwich based on SiC on the value of the magnetic field applied perpendicularly to the sample plane; the red color shows the calculation dependence of the oscillations; $I_{ds}$ = 10 nA, $T$ = 300K.

It should be noted that the area $S_{xx}$ of the edge channel between the *xx*-contacts $S_{xx}$ = 2000 $\mu$m × 1.54 nm = 3.08×10$^{-12}$ m$^2$ has $\frac{2000}{134}$ ≈15 single carrier, which corresponds to the value of the sheet density $n_{2D}$ ≈5×10$^{12}$ m$^{-2}$. This is in agreement with the values obtained from the field dependences of magnetic susceptibility [23]. The area $S_{xx}$ contributes to the obtained oscillations of magnetic susceptibility of the period $\Delta B_{13Oe}$ = 13Oe [23]: at the change of the magnetic field by this value, the magnetic flux through the area $S_{xx}$ increases by the value of the flux quantum $\Delta B_{13 E} \times S_{xx} = \Phi_0 = h/e$.

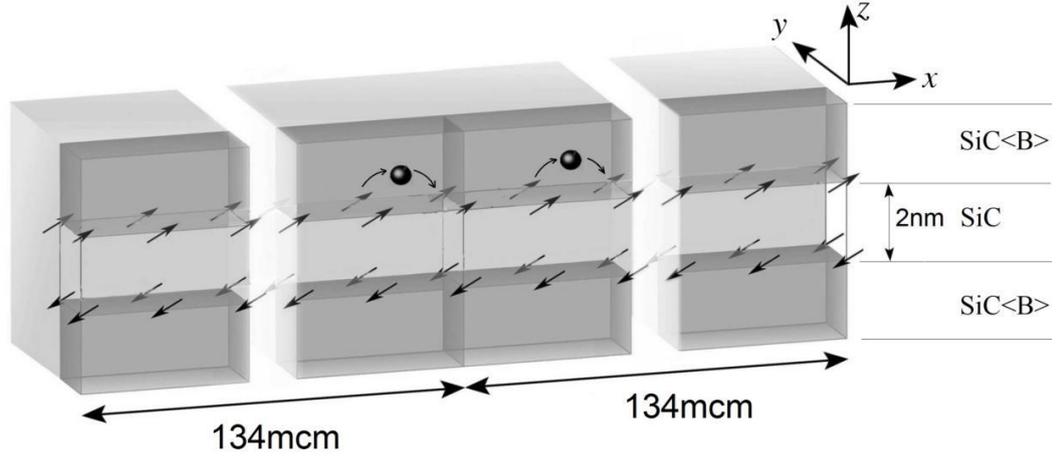

**Figure 4.** Scheme of regions of interference (pixels) of the single carriers with dimensions 134 $\mu$m × 1.54 $\mu$m × 2 nm. There is tunneling of the single carriers along a shell of boron dipoles, which cover the pixel.

Such regions-pixels exhibit a property of quantum conductivity [26]. That is why the edge channel can be considered as a ballistic one, wherein each of the pixels can be characterized by the resistance equal to the resistance quantum $h/(e^2)$. Furthermore, since the electron-electron interaction may be suppressed strongly under high pressures of several hundred GPa at the silicon substrate-silicone carbide interface, it is possible to have interference regions containing a pair of carriers with possible realization of Josephson junction near the microdefect interface. These conditions form the pixels of a double length with geometrical sizes 268 $\mu$m × 1.54 nm × 2 nm (Fig. 5), which are characterized by the resistance of $h/(4e^2)$.

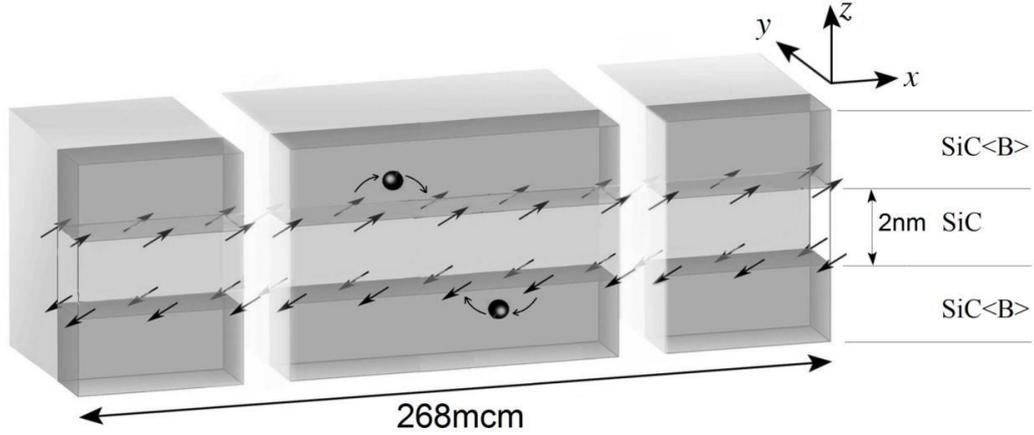

**Figure 5.** Scheme of the double-length pixel containing the pair of carriers and having the sizes 268 $\mu$m × 1.54 nm × 2 nm. There is tunneling of the pair carrier along a shell of boron dipoles, which covers the pixel.

The process of radiation generation can be evaluated. The longitudinal current $I_{ds}$ applied to the structure creates the magnetic field $B$. This field can be estimated by taking into account that $I_{ds}$ can form loops. The model of the current loop gives

$$B \approx \mu_0 \frac{I_{ds}}{2r}, \qquad (3)$$

where the effective radius $r$ is related by $r = \sqrt{S/\pi}$ to the area $S$ covered by the circuit.

Thus, the longitydinal current applied creates the flux $\Phi_0 = BS$ at the area $S$ and the flux change $\Delta\Phi$ in the individual double-length pixel inside this area; this process inside it also results in the generation current:

$$I_{gen} = \frac{U_{xy}}{R_{pixel}}, \qquad (4)$$

where $R_{pixel} = h/(4e^2)$ for the double-length pixel, and the value $Uxy$ is voltage measured at the $xy$-contacts of the structure depending on $Ids$. The results of these measurements are shown on Fig. 6, and it follows therefrom that $U_{xy} = 3.5\text{Ohm} \times I_{ds}$.

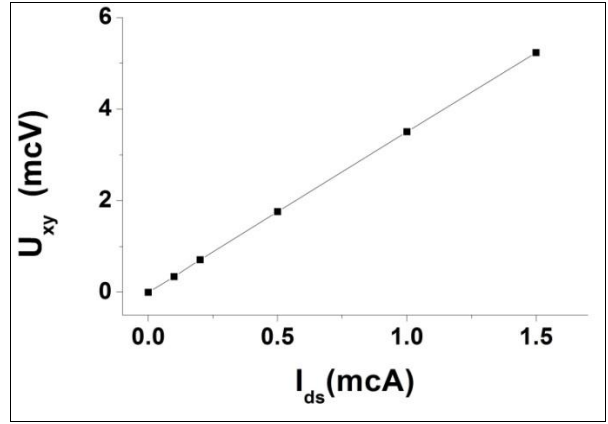

**Figure 6.** Dependence of the voltage $Uxy$ of the SiC-based nanosandwich on the longitudinal current $Ids$; $T = 300K$.

The frequency of emitted radiation can be found in accordance with the Faraday formula:
$$I_{gen} = \frac{\Delta E}{\Delta \Phi} = \frac{h\nu}{\Delta \Phi} \qquad (5)$$

The change of the flux in the individual double-length pixel
$$\Delta \Phi = \frac{m}{N}\Phi_0 \qquad (6)$$

depends on their number $N = S/(2S_{pixel})$ inside the area $S$ and the number of $m$ of the flux quantums $\Phi_0 = \frac{h}{2e}$, which are captured on the $N$ pixels; $m$ accepts the values $1, 2,\ldots, \frac{BS}{h/(2e)}$, the latter thereof corresponds to uniform population. Thus the frequency is equal to
$$\nu = I_{gen}\frac{m}{N}\frac{1}{2e}. \qquad (7)$$

When $I_{ds} = 30\text{mA}$, then we obtain $I_{gen} = 16.27\ \mu\text{A}$. If the loop of $Ids$ covers the area of the edge channel $S_{ds} = 4200\ \mu\text{m} \times 1.54\ \text{nm} = 6.5 \times 10^{-12}\ \text{m}_2$, then $\frac{m}{N} = \frac{41}{15}$ at the uniform population. At this, the frequency value at the minimum change of the flux: $\nu\left(\frac{m}{N} = \frac{1}{15}\right) = 3.389$ THz. If the loop of $I_{ds}$ covers the area of the whole sample $S_{device} = 4200\ \mu\text{m} \times 200\ \mu\text{m} = 8.4 \cdot 10^{-7}\ \text{m}^2$, then in these conditions $\frac{m}{N} \approx \frac{3}{412}$ at the homogeneously occupation. The frequency value at the minimum change of the flux: $\nu\left(\frac{m}{N} = \frac{1}{412}\right) = = 0.123$ THz. The obtained values are in good agreement with the values of 3.4 and 0.12 THz, which are found in detail study of features of the obtained spectra (see below).

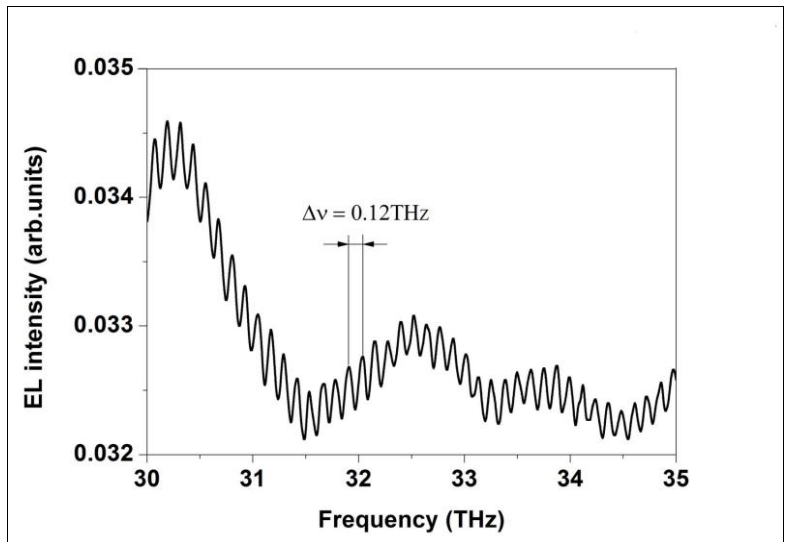

**Figure 7.** Fabry-Perot oscillations within the electroluminescence spectrum of the studied sample corresponding to the modulation frequency of 0.12 THz; $Ids = 30$ mA, $T = 300$ K.

Thus, in the mid- (up to 27 μm) and far- (up to 333 μm) infrared ranges of the wavelengths the spectra have modulations. In the framework of the Fabry–Perot resonator the distance $\Delta\nu$ between the adjacent peaks is related to the geometrical length $L_0$ of the resonator [27]:

$$\Delta\nu = \frac{c}{2L_0 n}, \qquad (8)$$

where $c$ — the speed of light in vacuum, $n$ — the refractive index of the medium. The found features correspond to the values of the modulation frequencies $\Delta\nu = 0.12$ THz (Fig. 7) and $\Delta\nu = 3.4$ THz (Fig. 8). At the value of $n = 2.55$ for silicon carbide (the database NSM [28]) the value $\Delta\nu = 0.12$ THz corresponds to the resonator of the length $L_0 = 490.0$ μm, while the value $\Delta\nu = 3.4$ THz corresponds to the resonator of the length $L_0 = 17.3$ μm.

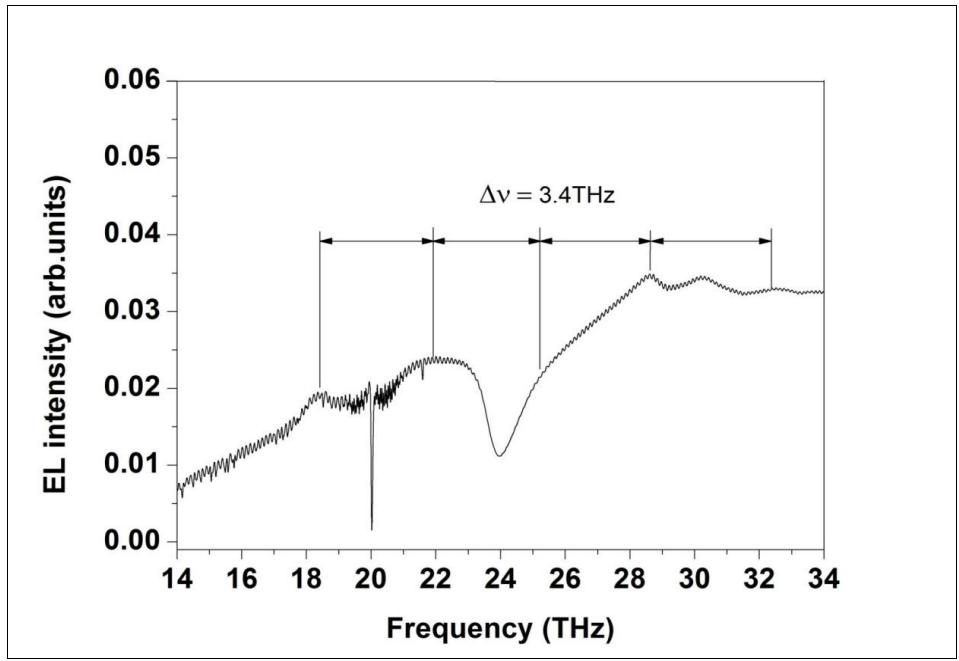

**Figure 8.** Fabry-Perot oscillations within the electroluminescence spectrum of the studied sample corresponding to the modulation frequency of 3.4 THz; $I_{ds} = 30$ mA, $T = 300$ K. Th dip at the frequency of 20 THz is correlated to the $CO_2$ absorption lines.

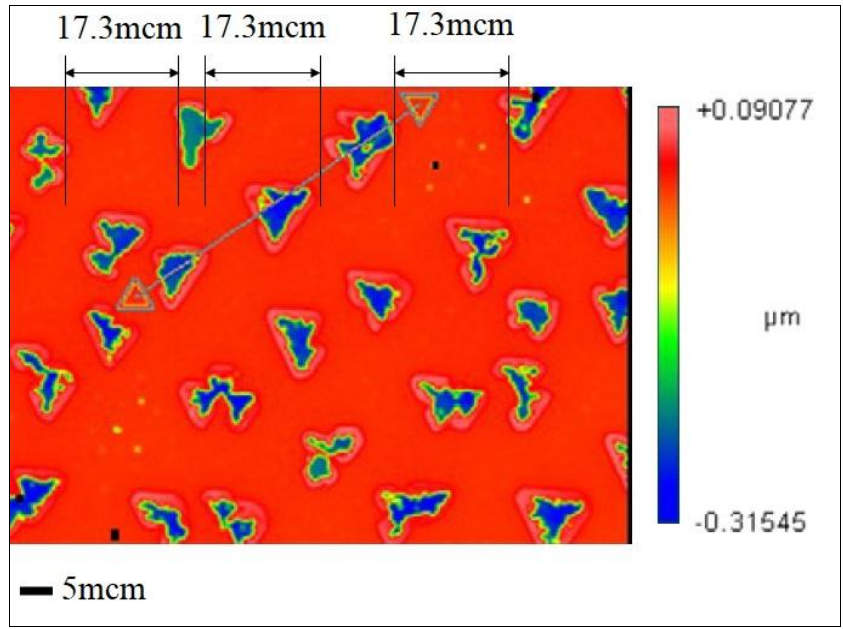

**Figure 9.** Image of the SiC surface on Si in film growth onditions at the CO low pressure and without SiH$_4$. The microdefects form systems of microresonators of the length of 17.3 μm, which correspond to the frequency of 3.4 THz. The value of the coordinate *z* in μm is shown in color (as per the data of the study [18]).

Defects separated by such a typical micron distance appear spontaneously at dislocations and other defects of the silicon substrate in the conditions of the reaction (1) at the CO low pressures [18]. At the same time, the surface of the SiC being formed is exhibiting open etching pits outgoing to the substrate surface, which remind "volcano craters", through which SiO gas is removed (out of the substrate bulk). This is shown on Fig. 9, which depicts the image of the SiC on Si, shot using the New View-6000 profile meter (produced by Zygo); the bigger values along the coordinate *z* are marked with the light color, so are the lower with the dark color. With increase in the CO pressure to the required value and addition to the SiH$_4$ system, the SiC nucleation rates become comparable on smooth sections of the surface and on the dislocations. As a result, the film surface becomes smooth. However, such microdefects can appear on edges of the Hall structure (Fig. 1) during doping in conditions of the non-equilibrium diffusion from the gas phase in the excessive flux of silicon vacancies from the surface. As a result, the edge channel forms the system of incorporated resonators, which correspond to the frequency $\Delta v$ = 3.4 THz.

The resonators with a typical millimeter size are in the structure after formation of the contact areas: the frequency $\Delta v$ = 0.12 THz corresponds to the regions with the size $f/2 \sim 0.5$mm between the vertical contacts (see Fig. 1).

## 4. Conclusion

Thus, it has found and investigated electroluminescence from the silicon carbide nanostructures, which are produced by the method of matched atom substitution. The electroluminescence spectra were recorded in the mid- and far-infrared ranges by means of the Bruker Vertex 70 Fourier spectrometer at the room temperature in conditions of drain-source passing of the longitudinal current in the Hall geometry structures. The amplitude and frequency modulation of the obtained spectra is found in the terahertz frequency range, 3.4 and 0.12 THz. The found terahertz radiation occurs due to the Faraday quantum effect in the conditions of captured single quantums of the magnetic flux inside the edge channels of the nanostructure. Within the suggested model of Faraday quantum effect, the single quanta of the magnetic flux induced during flowing of the drain-source current result in current generation in the edge channels and, respectively, to the terahertz radiation, whose frequency depends on the geometrical parameters of the studied nanostructures.

**Funding**

This study was carried out under financial support of the RSF grant (the grant No. 20-12-00193).

**Acknowledgments**

The synthesis of a SiC layer on Si was performed using the equipment of the "Physics, Chemistry, and Mechanics of Chips and Thin Films" unique scientific unit at the Institute of Problems of Mechanical Engineering, Russian Academy of Sciences (St. Petersburg).

**Conflict of interest**

The authors declare that they have no conflict of interest.

**References**

[1] M. Danciu, T. Alexa-Stratulat, C. Stefanescu, G. Dodi, B.I. Tamba, C. Teodor Mihai, G.D. Stanciu, A. Luca, I.A. Spiridon, L.B. Ungureanu, V. Ianole, I. Ciortescu, C. Mihai, G. Stefanescu, I. Chirila, R. Ciobanu, V.L. Drug. Materials, **12** (9), 1519 (2019).
[2] S. Zhong. Front. Mech. Eng., **14** (3), 273 (2019).
[3] A.K. Panwar, A. Singh, A. Kumar, H. Kim. IJET-IJENS, **13**, 33 (2013).
[4] R. Lewis. J. Phys. D: Appl. Phys., **47** (37), 374001 (2014).
[5] L. Consolino, S. Bartalini, P. De Natale. J. Infr. Milli Terahz Waves, **38**, 1289 (2017).
[6] E.M. Gershenzon, M.B. Golant, A.A. Negirev, V.S. Savel'ev. *Lampy obratnoy volny millimetrovogo i submillimetrovogo diapazonov voln,* pod red. N.D. Devyatkova (M., Radio i svyaz', 1985) (in Russian).
[7] Q. Sun, Y. He, E.P.J. Parrott, E.P. MacPherson. J. Biophotonics, **11** (2), e201700111 (2018).
[8] https://www.menlosystems.com/products/thz-time-domain-solutions/ terak15-terahertz-spectrometer/
[9] R. Köhler, A. Tredicucci, F. Beltram, H.E. Beere, E.H. Linfield, A.G. Davies, D.A. Ritchie, R.C. Iotti, F. Rossi. Nature, **417**, 156 (2002).
[10] J. Faist, F. Capasso, D.L. Sivco, A.L. Hutchinson, A.Y. Cho. Science, **264**, 553 (1994).
[11] L. Ozyuzer, A.E. Koshelev, C. Kurter, N. Gopalsami, Q. Li, M. Tachiki, K. Kadowaki, T. Yamamoto, H. Minami, H. Yamaguchi, T. Tachiki, K.E. Gray, W.-K. Kwok, U. Welp. Science, **318**, 1291 (2007).
[12] N.T. Bagraev, P.A. Golovin, V.S. Khromov, L.E. Klyachkin, A.M. Malyarenko, V.A. Mashkov, B.A. Novikov, A.P. Presnukhina, A.S. Reukov, K.B. Taranets. J Altern. Complement Integr. Med., **6**, 112 (2020). Semiconductors, 2022.